\documentclass[twocolumn]{aastex631}
\usepackage{amssymb, amsmath}
\usepackage{natbib}
\usepackage{color}
\usepackage{ulem}
\usepackage{appendix}
\usepackage{hyperref}
\usepackage{relsize}
\usepackage{mathtools}

\definecolor{red}{rgb}{1.0,0.0,0.0}

\begin{document}

\title{Direct Imaging Detection of the Protoplanet AB Aurigae b at Wavelengths Covering Pa$\beta$: \\Rebuttal to Biddle et al. (2024)}


\author{Thayne Currie}
\affiliation{Department of Physics and Astronomy, University of Texas-San Antonio, San Antonio, TX, USA}
\affiliation{Subaru Telescope, National Astronomical Observatory of Japan, 
650 North A`oh$\bar{o}$k$\bar{u}$ Place, Hilo, HI USA}
\email{thayne.currie@utsa.edu}
\shortauthors{Currie}
\begin{abstract}
Recently, Biddle et al. (2024) claimed a non-detection of the protoplanet AB Aurigae b in Keck/NIRC2 Pa$\beta$ imaging.  I reprocess these newly-public data and compare them to data from the extreme AO platform (SCExAO/CHARIS) used to discover AB Aur b.  AB Aur b is decisively imaged with SCExAO/CHARIS at wavelengths covering Pa$\beta$.  The Biddle et al. non detection of AB Aur b results from a far poorer image quality that is non competitive with SCExAO/CHARIS.  Their contrast limits and thus constraints on accretion are overestimated due to an inaccurate AB Aur b source model.  Consequentially, the revised Pa$\beta$ 2-$\sigma$ upper limit from these data is about three times higher than previously reported.   Irrespective of image quality, single-band Pa$\beta$ imaging is ill suited to conclusively identifying accretion onto AB Aur b.  Instead, high-resolution H$\alpha$ spectroscopy may provide accretion signatures.
Aside from PDS 70, AB Aurigae remains the system with the strongest evidence for having a directly-imaged protoplanet. 
\end{abstract}

\section{Introduction}
\textit{Protoplanets} directly imaged within their natal protoplanetary disks shed light on the formation of fully-formed jovian exoplanets \citep[][]{Keppler2018,Haffert2019,Benisty2023,Currie2023b}.  The young star AB Aurigae hosts a spatially-extended object identified as an embedded protoplanet at a wide ($\sim$90 au) separation \citep{Currie2022ABAurb} (hereafter C22) primarily based on its near-infrared (IR) spectrum, low polarization, and counterclockwise orbital motion.  C22 also detected AB Aur b in H$\alpha$, but the source of this emission -- accretion or scattered light -- was deemed unclear.  If AB Aur b has detectable accretion, it may have identifiable emission in other lines like H$\beta$ or Pa$\beta$ \citep[][]{Eriksson2020,Demars2023}.




Recently, \citet{Biddle2024} (hereafter B24) claimed a non-detection of AB Aur b in Pa$\beta$ imaging based on a Keck/NIRC2 imaging sequence using conventional adaptive optics.  They suggest that their analysis places new constraints on accretion onto AB Aur b from an upper limit on Pa$\beta$ line emission: namely, that their data show AB Aur b to either be weakly accreting or have no accretion at all.

To assess these results, I reprocess these now-public data and compare results to those from the 
ground-based platform used to discover AB Aur b (C22): the
\textit{Subaru Coronagraphic Extreme Adaptive Optics} Project coupled with the CHARIS integral field spectrograph \citep{Jovanovic2015,Groff2016} (SCExAO/CHARIS).  
My analysis reaffirms AB Aur b's detection at wavelengths covering Pa$\beta$, revises results from the B24 data, and assesses whether Pa$\beta$ imaging can effectively identify accretion.
\section{Data and Analysis}

 

I consider the October 2 2020 SCExAO/CHARIS observations, the most recent and one of the highest-quality SCExAO/CHARIS data sets presented in C22.  I reduced these using the CHARIS data processing pipeline as in C22.   The CHARIS point-spread function (PSF) as seen from CHARIS's satellite spots 
shows sharp Airy rings in each spectral channel.
The PSF appears diffraction limited, as expected for a system that achieves near-IR Strehl ratios of 0.70-0.94.  In the highest-quality exposures, the AB Aur disk and AB Aur b are faintly visible in raw, 31-second channel-combined CHARIS images.  For PSF subtraction, I followed the conservative, full-frame A-LOCI approach described in C22 in combination with polarimetry-constrained reference star differential imaging (PCRDI) also presented in C22 and described in full in \citet{Lawson2022}.


\begin{figure*}
    \centering
    \includegraphics[width=0.495\textwidth,clip]{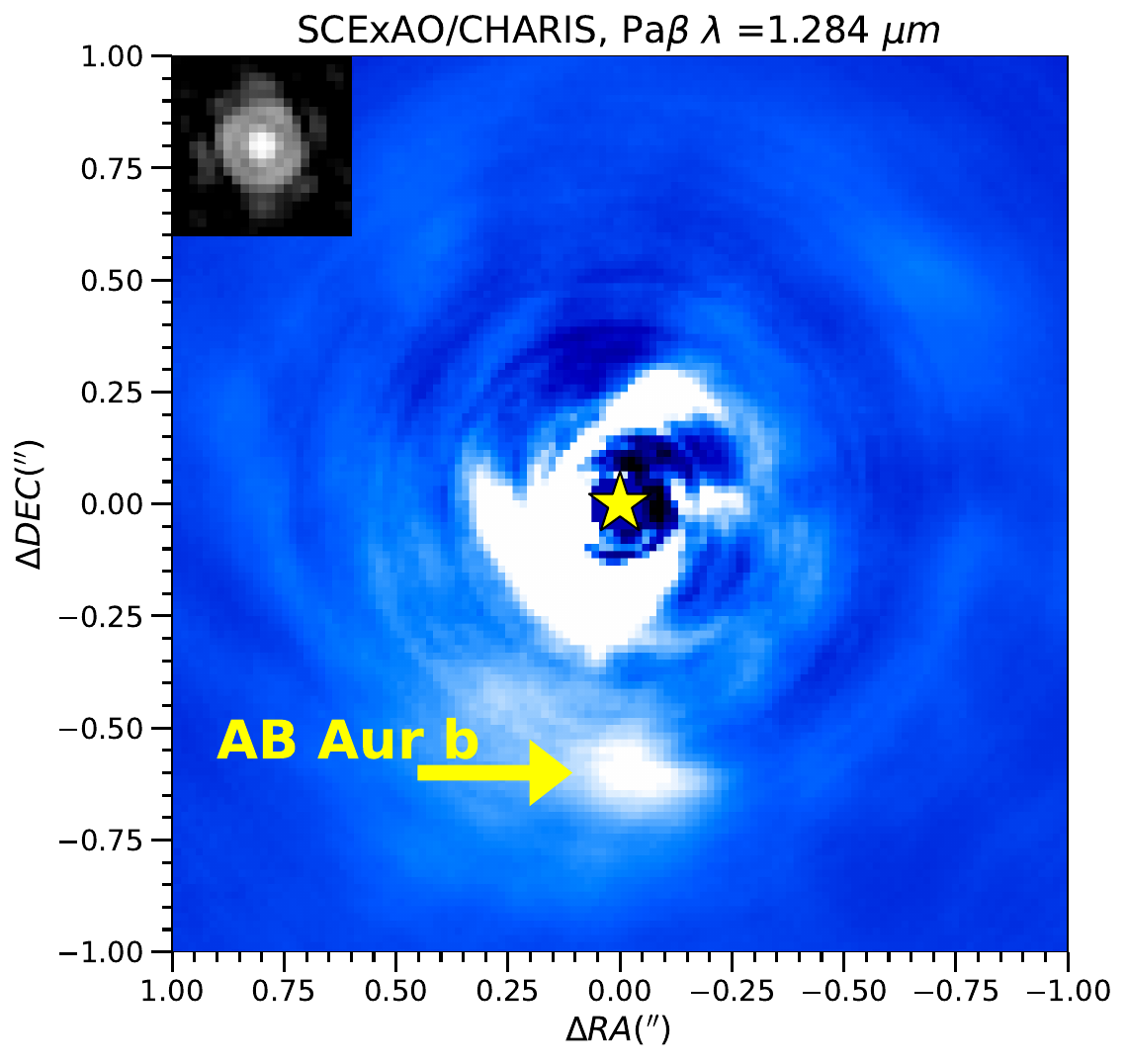}
    \includegraphics[width=0.495\textwidth,clip]{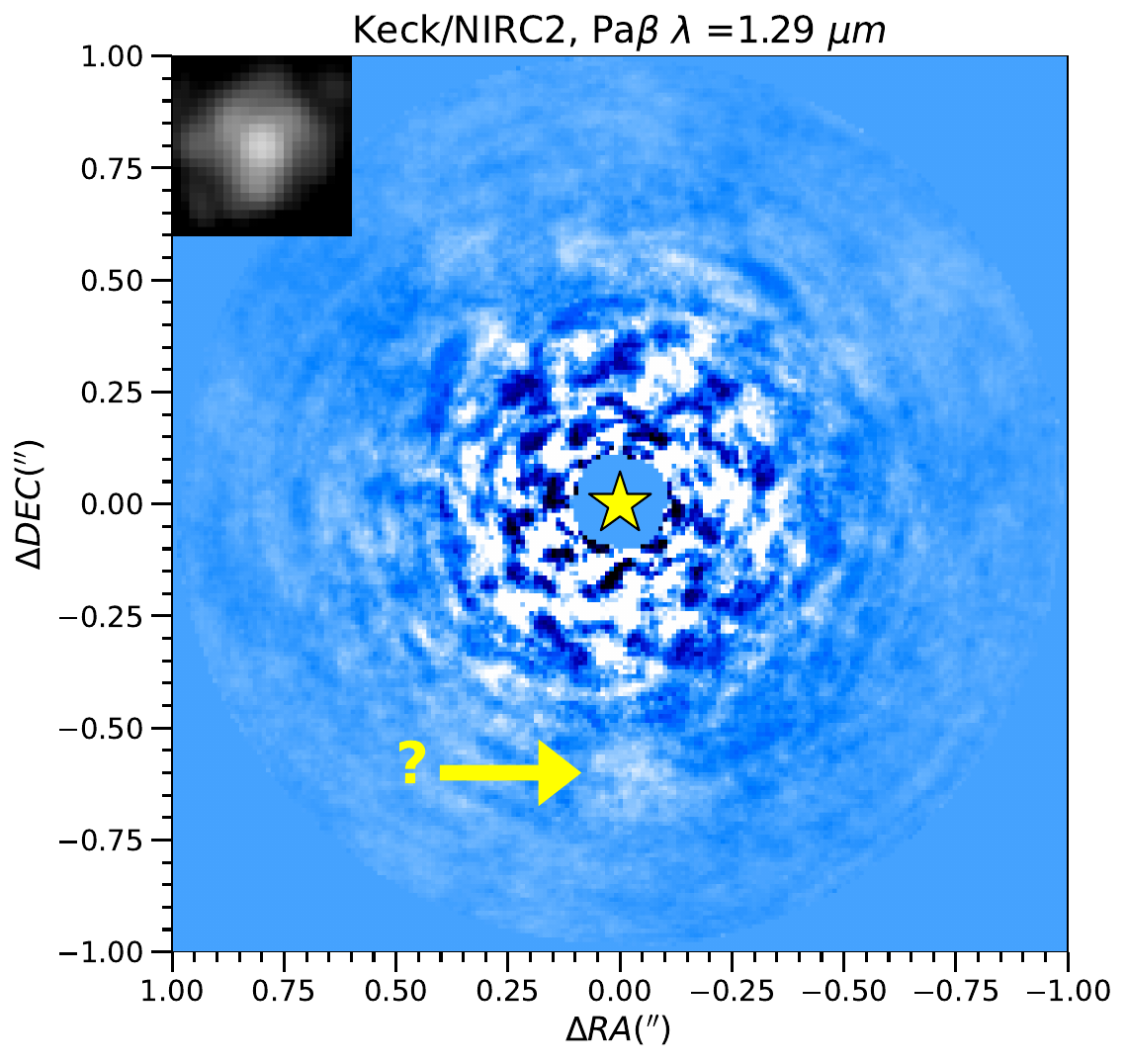}
    \caption{SCExAO/CHARIS (left) and reprocessed Keck/NIRC2 images (right) of AB Aurigae: insets show the normalized instrumental PSFs at Pa$\beta$ (identical intensity scaling).
  SCExAO/CHARIS detects AB Aur b in Pa$\beta$; my reprocessed Keck/NIRC2 image shows a nondetection or possibly low-significance detection of AB Aur b. 
   }
    
    \label{carefulrefereeingisgood}
\end{figure*}

I downloaded the B24 AB Aur Pa$\beta$ data ($\lambda$ $\sim$ 1.29 $\mu m$; $\Delta$$\lambda$ $\sim$ 0.02$\mu m$) from the \textit{Keck Observatory Archive} (Program ID: 2022A$\_$N064; PI. B. Bowler).  Subsequent data reduction steps utilized a well-tested pipeline \citep{Currie2011,Currie2023a}.  For PSF subtraction, I adopted the same conservative, full-frame implementation of A-LOCI shown to detect AB Aur b in C22, varying reduction parameters (e.g. the rotation gap, $\delta$, optimization area $N_{\rm A}$, and size of the reference library).  Unsaturated images revealed a PSF lacking Airy rings and containing a bright and variable speckle halo.  
The azimuthally-averaged full-width at half maximum (FWHM) exceeds 0\farcs{}041 for each unsaturated exposure,
far from diffraction-limited performance ($\theta_{\rm FWHM}$ $\approx$ 0\farcs{}027).  
The average Strehl ratio -- determined from the observatory-supplied function \texttt{nirc2strehl.pro}, modified for a revised pixel scale of 0\farcs{}009971 \citep{Service2016} --  is 0.13.

Figure \ref{carefulrefereeingisgood} shows the resulting images.  
AB Aur b is decisively detected in CHARIS channel 4 ($\lambda$ $\sim$ 1.284 $\mu m$; $\Delta$$\lambda$ $\sim$ 0.04$\mu m$) covering Pa$\beta$ (top-left panel).  
The AB Aur protoplanetary disk is likewise resolved down to the coronagraph mask edge (0\farcs{}13).  
In the reduction shown, AB Aur b's contrast with respect to the star in the CHARIS Pa$\beta$ channel is $\Delta$Pa$\beta$ $\sim$ 9.53.  As is clear from inspection of C22, AB Aur b is detected in Pa$\beta$ in multiple alternate reductions at different epochs: the median Pa$\beta$ contrast from all spectra shown in C22 is 9.42\footnote{C22 reports that the characteristic signal-to-noise ratios (SNR) of the detections are $\sim$10-12 in the highest quality data.  However, this calculation is highly conservative, as it computes the azimuthally-averaged noise from the final, sequence-combined, north-aligned images, which preserves highly-structured disk signals.  The disk signals then enter into the calculation of the noise, even though they are not residual speckles. Sequence-combined images derotated the opposite way from north-up smears out these signals, and better represents the true residual noise floor: the SNR of AB Aur b in this case would be $\sim$ 30 in the October 2020 data.}.

Keck/NIRC2 fails to yield a $\ge$5$\sigma$ detection of any astrophysical signal (right panel).   A subset of reductions yield an extended signal at [E,N] $\sim$ [-0\farcs{}04, -0\farcs{}60] consistent with the morphology, brightness, and latest position of AB Aur b (Currie et al. 2024 in prep.).  But it is only $\sim$2--2.5$\sigma$ significant.


I computed the Keck/NIRC2 5-$\sigma$ contrast limits from the azimuthally-averaged radial noise profile \citep{Currie2011,Currie2023a}.  Following standard practices, the noise at each radius is determined after replacing each pixel with its sum within a FWHM-sized aperture \citep[e.g.][]{Currie2011,Mawet2014} and computing the robust standard deviation (e.g. using Tukey's biweight theorem) of summed pixel values\footnote{If instead I sum within a 2$\times$FWHM-sized aperture as in B24, the resulting contrast curve is similar, at most $\sim$10-20\% fainter for extended sources at some separations considered here.}.    
I corrected the contrast curves 
for signal loss due to self-subtraction/oversubtraction
via forward-modeling for a point source and extended source model appropriate for AB Aur b (C22).   Unsaturated Pa$\beta$ images provide the NIRC2 instrumental PSF.  For the AB Aur b model, I convolved its intrinsic intensity distribution -- a gaussian with a Half-Width at Half-Maximum (HWHM) of 0\farcs{}045 (C22) -- with the NIRC2 instrumental PSF, yielding an extended source of $\theta_{\rm FWHM}$ $\sim$ 0\farcs{}12.  
Injecting synthetic sources into the data and recovering them after PSF subtraction confirmed my limits.

 \begin{figure*}
    \centering
    \includegraphics[width=0.495\textwidth,clip]{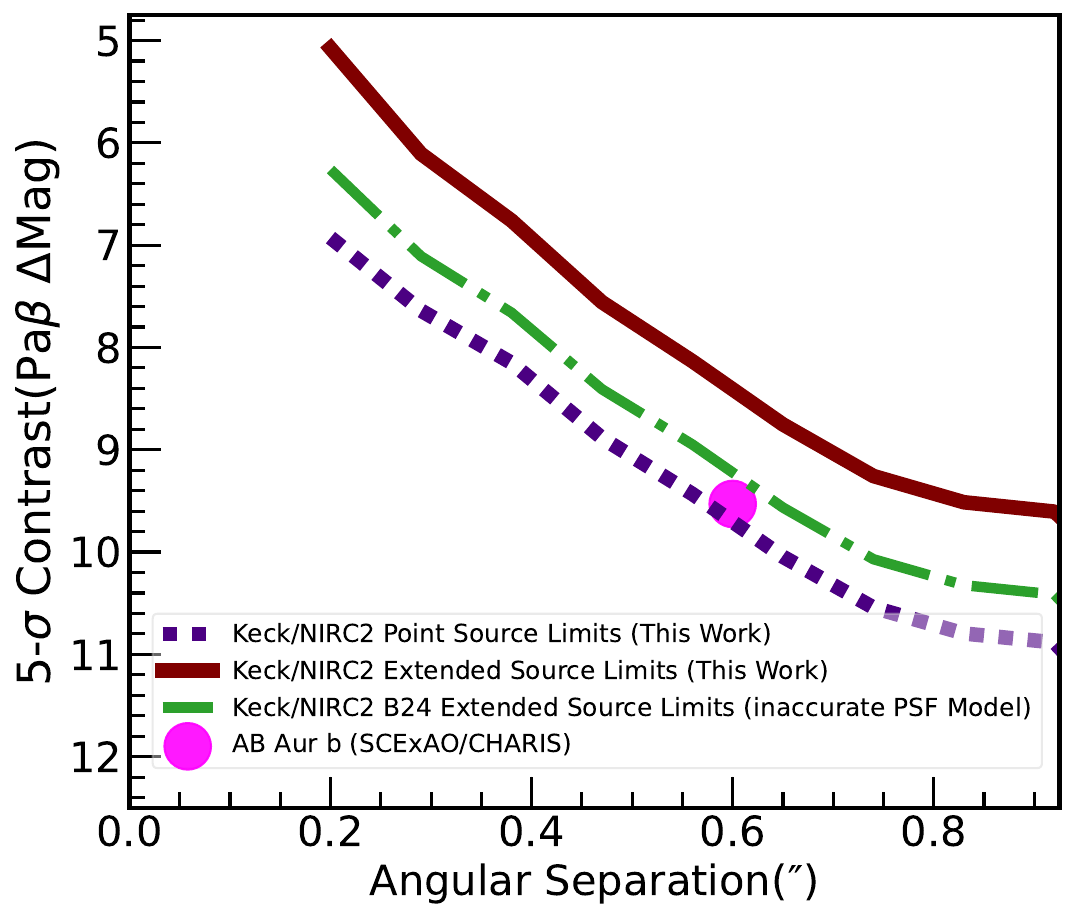}
    \includegraphics[width=0.495\textwidth,clip]{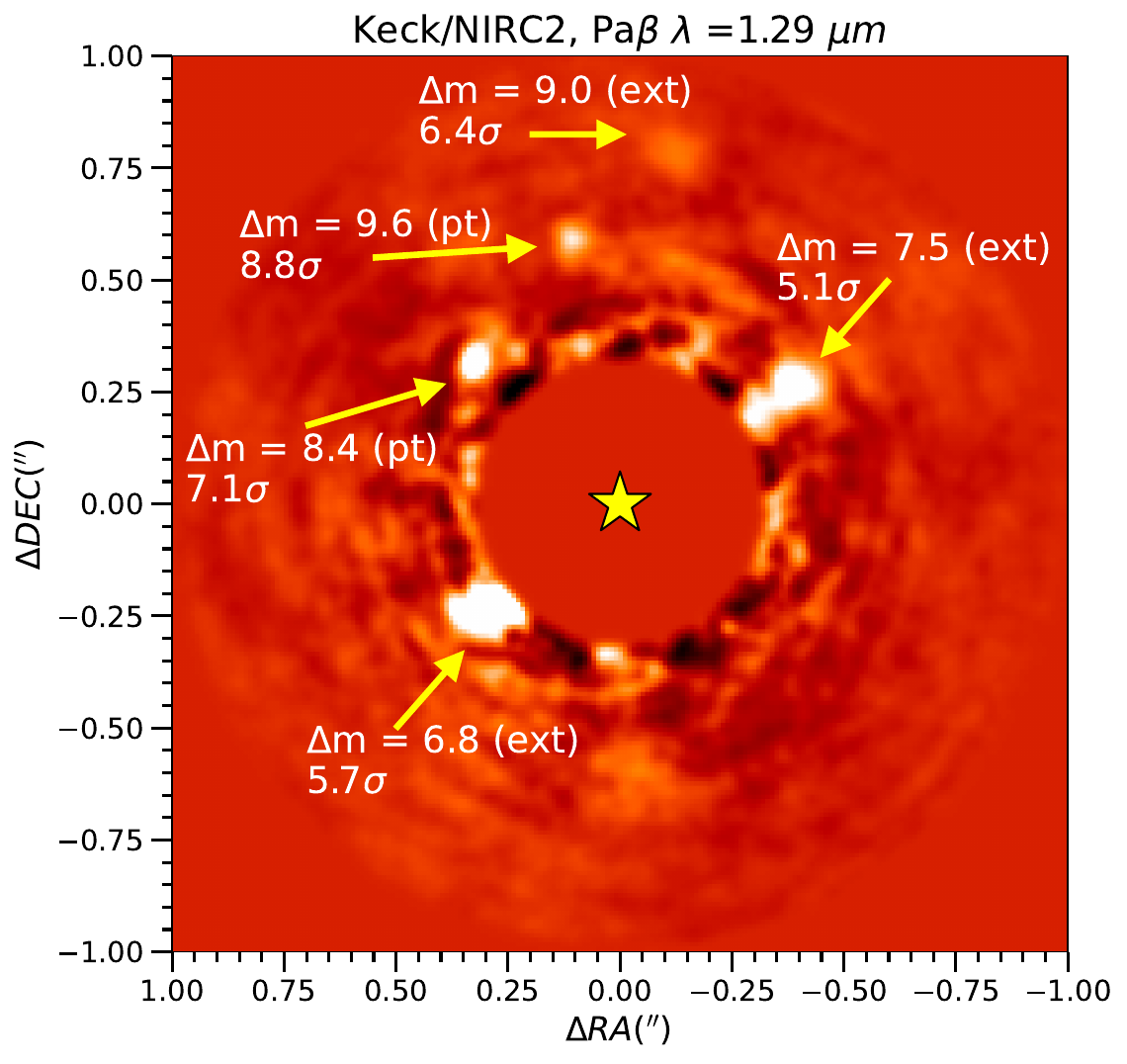}
    \caption{
    (left) My Keck/NIRC2 contrast curve for point sources, an accurate extended source model, and the inaccurate used in B24.  (right) Keck/NIRC2 image with injected point sources (pt) and extended sources (ext). 
    For clarity, I mask the image's inner 0\farcs{}3 and smooth it by a FWHM-sized gaussian kernel.}
    
    \label{carefulrefereeingisgood2}
\end{figure*}

At 0\farcs{6}, my point source contrast limits equal those from B24: at $\rho$ $\sim$ 0\farcs{}2--0\farcs{}3, they are $\sim$0.5 mags deeper  (Figure \ref{carefulrefereeingisgood2}, left panel).   However, contrast limits for an AB Aur b-like source are 1.2 magnitudes shallower than claimed before
($\Delta$Pa$\beta$ $\sim$ 8.4).  
Injected extended sources at $\rho$ $\approx$ 0\farcs{}5--0\farcs{}7 equal to B24's stated 5-$\sigma$ limits are only $\sim$1.5--2$\sigma$ significant and not clearly distinguishable from residual speckle noise: 5$\sigma$-significant sources are much brighter than their limits (right panel).


Despite agreement for point source limits, my extended source contrast limits differ from B24's because they adopt an inaccurate AB Aur b source model.   First, their assumed gaussian point-source intensity profile poorly matches the empirical NIRC2 PSF. At aperture radii of 0.5 $\lambda$/D--1 $\lambda$/D, the empirical PSF's encircled energy is 1.6--1.8 times lower than the gaussian model's encircled energy.

Second, their model for AB Aur b's intrinsic intensity distribution -- a flat-intensity, 0\farcs{}045-radius filled circle -- is erroneous.  C22's 0\farcs{}045-radius source size instead refers to the source distributions's \textit{HWHM}\footnote{See paragraph 3 in the \textit{Source astrometry, photometry and spectroscopy: ground-based data} section of the Supplementary Materials in C22.   
As another check on accuracy, their AB Aur b model convolved with their instrumental PSF model (a gaussian intensity profile) has a \textit{smaller} FWHM than the intrinsic (deconvolved) source distribution estimate listed in C22.  Similarly, as can be shown using either IDL or Python (e.g. \texttt{psf$\_$gaussian.pro} and \texttt{fullwid$\_$halfmax.pro} or the \texttt{Gaussian2D} astropy.modeling and \texttt{convolve2d} scipy.signal classes) a 0\farcs{}045-radius constant-intensity circle convolved with the CHARIS PSF cannot reproduce the observed CHARIS intensity distribution of AB Aur b.}.   
The correct AB Aur b model convolved with the empirical PSF has a normalized peak intensity and encircled energy at  0.5 $\lambda$/D--1 $\lambda$/D 2.1--2.2 times lower than the B24 model.  Using the B24 source model, I obtain contrast limits similar to theirs at 0\farcs{}6. 
However, 
AB Aur b is more diffuse than in their model: its per-pixel signal is fainter and its signal loss from self-subtraction is greater, leading to shallower contrast limits.  
As a result, 2-$\sigma$ line flux density upper limit from B24 data is now $\sim$ 9.44$\times$10$^{-15}$ W m$^{-2}$ $\mu m^{-1}$, a factor of three higher than previously reported.

\section{Discussion}
In this work, I reprocessed the Keck/NIRC2 Pa$\beta$ data presented in B24 and compare them to SCExAO/CHARIS data presented from the AB Aur b discovery paper \citep{Currie2022ABAurb}.   
AB Aur b is detected with SCExAO/CHARIS at wavelengths covering Pa$\beta$.  In contrast, the Keck/NIRC2 data presented by B24 are too shallow to yield a 5-$\sigma$ detection, although my rereduction may identify it at a low SNR.  SCExAO/CHARIS's superior image quality explains the difference between its detection and the Keck/NIRC2 results\footnote{ The SCExAO/CHARIS data's longer integration time (67 minutes vs. 16.7 minutes) and greater parallactic angle rotation (127$^{o}$ vs. 57.9$^{o}$) further increase its competitive difference with the B24 data.}.
An inaccurate source intensity model explains B24's Keck/NIRC2 5-$\sigma$ contrast limits for an AB Aur b-like sources\footnote{B24 also state that a tension may exist between their Pa$\beta$ contrast limits and the C22 contrast of $\Delta$J $\sim$ 9.2.  Irrespective of revisions of the Keck/NIRC2 contrast limits described in this work that eliminate this tension, this claimed tension is largely alleviated by noting that the Keck/NIRC2 Pa$\beta$ filter is a narrowband filter ($\lambda$ $\sim$ 1.28 $\mu m$, $\Delta$$\lambda$ $\sim$ 0.02 $\mu m$), while J band is different, referring to the standard broadband filter ($\lambda$ $\sim$ 1.248 $\mu m$, $\Delta$$\lambda$ $\sim$ 0.163 $\mu m$).  The estimated Pa$\beta$ contrasts from CHARIS reported in this paper and drawn from data in C22 are roughly 25\%--35\% steeper than the J band contrast.  To explain the Keck/NIRC2 Pa$\beta$ non-detection, the paper also hypothesizes that the CHARIS absolute flux calibration may be inaccurate.  However, the flux calibration using satellite spots has been determined using both internal source data from two independent measurements and on-sky data and shows agreement to within $\sim$8\% across epochs \citep{Currie2020b,Chen2023}.  Additionally, AB Aur b photometry from HST/NICMOS presented in C22 is in full agreement with CHARIS measurements.}.

Irrespective of image quality, the B24 data and any near-term Keck/NIRC2 single-band Pa$\beta$ imaging are ill suited to conclusively identifying accretion onto AB Aur b. 
Pa$\beta$ line emission is measured \textit{spectroscopically}.  Modeled narrowband \textit{photometry} in B24 poorly constrains the star's Pa$\beta$ emission contemporaneous with imaging (EW = -94 $\pm$ 64 $\dot{A}$).  At the 1.5$\sigma$ level, the star could have strong Pa$\beta$ emission or none at all, precluding an identification of differences between its Pa$\beta$ emission and AB Aur b's.

Besides accretion, Pa$\beta$ signal from AB Aur b's position can originate from thermal emission and scattered starlight, since AB Aur b is an additive signal to the disk (C22).   Distinguishing these three emission sources requires a precise calibration of the background disk signal -- with self-subtraction mitigated via forward-modeling -- and an estimate of the continuum level (thermal emission) at flanking spectral channels.  By definition, this is not possible for \textit{single-band} Pa$\beta$ imaging.
Even with SCExAO/CHARIS's image quality and adopting PSF subtraction approaches resulting in no self-subtraction (RDI) or even little signal loss at all (PCRDI) as in C22\footnote{The reduction approach used in B24 -- ADI/KLIP \citep{Soummer2012-KLIP} in narrow annuli or small wedge-like sectors -- is highly inadvisible for the detection and accurate morphological interpretation of protoplanets in highly structured disks.  Used in this manner, KLIP acts as a high-pass filter of the combined planet + disk signal.  Its usage with ADI results in the attenuation of the protoplanet signal via self-subtraction \textit{and} subtraction of the protoplanet signal by the disk.  Calibrating this effect and recovering the true combined signal is exceptionally challenging and requires forward-modeling of both the planet and disk signal.}, the extracted spectra at Pa$\beta$ show uncertainties on the 10-20\% level due to fundamental challenges with measuring and extracting the AB Aur b signal from the disk background (see C22).

Finally, some accreting planet-mass objects show Pa$\beta$ line emission \citep{Demars2023}.  However, the PDS 70 bc \textit{protoplanets} have Pa$\beta$ non detections thus far: line flux upper limits are potentially in conflict with model predictions drawn from the objects' H$\alpha$ emission \citep{Uyama2021}.   High-resolution H$\alpha$ spectroscopy -- e.g. VLT/MUSE -- presents the clearest path to identify accretion onto AB Aur b \citep[e.g.][]{Haffert2019,Hashimoto2020}.

AB Aur b shows evidence for orbital motion, has a low polarization signal, and has a near-IR spectrum that appears different from scattered starlight.   It has been detected in over twelve data sets spanning thirteen years with three telescopes using six instruments.  Aside from PDS 70, AB Aur remains the system with the strongest evidence for having an imaged protoplanet.  Constraints on accretion properties help to complete a picture of AB Aur b.
\\\\\\

I thank Christian Marois and Taichi Uyama for helpful comments on an earlier draft of this paper.

I acknowledge the very significant cultural role and reverence that the summit of Mauna Kea holds within the Hawaiian community.  I am most fortunate to have the opportunity to conduct observations from this mountain.

This research has made use of the Keck Observatory Archive (KOA), which is operated by the W. M. Keck Observatory and the NASA Exoplanet Science Institute (NExScI), under contract with the National Aeronautics and Space Administration.

\software{:\citep{numpy}, \texttt{Jupyter Notebooks} \citep{jupyter}, \texttt{Matplotlib} \citep{matplotlib},  \texttt{Astropy} \citep{astropy:2013,astropy:2018,astropy:2022}, \texttt{SciPy} \citep{scipy}, \texttt{CHARIS DRP} \citep{Brandt2017}, \texttt{CHARIS DPP} \citep{Currie2020b}.}

\bibliography{bibliography}{}
\bibliographystyle{aasjournal}


\end{document}